\documentclass[aps,prd,twocolumn,amsmath,showpacs,superscriptaddress,nofootinbib]{revtex4-1}
\usepackage{graphicx}
\usepackage{longtable}
\usepackage{float}
\usepackage{dcolumn}
\usepackage{bm}
\usepackage{appendix}

\newcommand{\Mpc}{\mathrm{Mpc}}

\begin{document}

%\preprint{}

\title{Featuring the primordial power spectrum: new constraints on interrupted slow-roll from CMB and LRG data}


\author{Micol Benetti}
\email[]{Micol.Benetti@roma1.infn.it}
%\thanks{}
\affiliation{Physics Department and ICRA, Universit\`a di Roma 
	``La Sapienza'', Ple.\ Aldo Moro 2, 00185, Rome, Italy}
\affiliation{Physics Department and INFN, Universit\`a di Roma 
	``La Sapienza'', Ple.\ Aldo Moro 2, 00185, Rome, Italy}
%%%%%
\author{Stefania Pandolfi}
%\thanks{}
\affiliation{Dark Cosmology Centre, Niels Bohr Istitute, University of Copnehagen, Juliane Maries Vej 30 , 2100 Copenhagen, Denmark}
%%%%%	
\author{Massimiliano Lattanzi}
\email[]{lattanzi@ferrara.infn.it}
%\thanks{}
%\altaffiliation{}}
\affiliation{Dipartimento di Fisica G. Occhialini, Universit\`a
Milano-Bicocca and INFN, sezione di Milano-Bicocca, Piazza della Scienza 3,
I-20126 Milano, Italy.}
\affiliation{Dipartimento di Fisica, Universit\`a di Ferrara and INFN, sezione di Ferrara, 
Polo Scientifico e Tecnologico - Edificio C Via Saragat, 1, I-44122 Ferrara Italy
}
\author{Matteo Martinelli}
\affiliation{SISSA, Via Bonomea 265, Trieste, 34136, Italy}
\affiliation{INFN, Sezione di Trieste, Via Valerio 2, 34127 Trieste, Italy}
\author{Alessandro Melchiorri}
%\thanks{}
\affiliation{Physics Department and INFN, Universit\`a di Roma 
	``La Sapienza'', Ple.\ Aldo Moro 2, 00185, Rome, Italy}

\date{\today}

\begin{abstract}
Using the most recent data from the WMAP, ACT and SPT experiments, we update the constraints on models with oscillatory features in the primordial power spectrum of scalar perturbations. This kind of features can appear in models of inflation where slow-roll is interrupted, like multifield models. We also derive constraints for the case in which, in addition to cosmic microwave observations, we also consider the data on the spectrum of luminous red galaxies from the 7th SDSS catalog, and the SNIa Union Compilation 2 data. We have found that: (i) considering a model with features in the primordial power spectrum increases the agreement with data with the respect of the featureless ``vanilla'' $\Lambda$CDM model  by $\Delta\chi^2 \simeq 7$; (ii) the uncertainty on the determination of the standard parameters is not degraded when features are included; (iii) the best fit for the features model locates the step in the primordial spectrum at a scale $k\simeq 0.005$ Mpc$^{-1}$, corresponding to the scale where the outliers in the WMAP7 data at $\ell=22$ and $\ell=40$ are located.; (iv) a distinct, albeit less statistically significant peak is present in the likelihood at smaller scales, with a $\Delta\chi^2 \simeq 3.5$, whose presence might be related to the WMAP7 preference for a negative value of the running of the scalar spectral index parameter; (v) the inclusion of the LRG-7 data do not change significantly the best fit model, but allows to better constrain the amplitude of the oscillations.

\end{abstract}

%\keywords{}

\maketitle

\section{Introduction}

The inflationary paradigm is an integral part of the currently accepted concordance cosmological model, explaining the flatness and homogeneity of
the observed Universe, as well as providing a mechanism to produce the primordial curvature perturbations that eventually led to the formation of structures. 
The shape of the power spectrum of primordial perturbations can be constrained, at least at the largest scales, using cosmic microwave background (CMB) data. 
The 7-year WMAP data are in excellent agreement with the assumption of a nearly scale-invariant power spectrum of scalar perturbations \cite{Komatsu:2010fb,Larson:2010gs}. Such a spectrum, described by a simple power law with spectral index $n_s$ very close to (albeit different from) unity, is the one that would be produced in the simplest inflationary scenario, that of a single, minimally-coupled scalar field slowly rolling down a smooth potential. The expectation of a power-law spectrum continues to hold up against scrutiny also when tested against observations at scales smaller than those probed by WMAP, like the small-scale CMB measurements of the Atacama Cosmology Telescope (ACT) \cite{Das:2010ga,Dunkley:2010ge,Hlozek:2011pc} and South Pole Telescope (SPT) ~\cite{Keisler:2011aw}, and the spectrum of luminous red galaxies ~\cite{beth}.
Nevertheless, a scale invariant power spectrum with $n_s=1$ could be easily put in agreement with data in some non-minimal models, e.g. considering an extended reionization process \cite{Mortonson:2008rx,Pandolfi:2010dz,Pandolfi:2010mv}, non standard processes during recombination like dark matter annihilation \cite{Galli:2009zc,Galli:2011rz,Giesen:2012rp,Evoli}, extra relativistic particles (see e.g. \cite{Hamann:2010pw,Archidiacono:2011gq}) and so on. 

In spite of this, however, models with localized ``features'' in the primordial power spectrum provide a better fit to the data \cite{Peiris:2003ff,Covi:2006ci,Hamann:2007pa,Mortonson:2009qv,Hazra:2010ve,Benetti:2011rp,Meerburg:2011gd} with respect to a smooth power-law spectrum. This is mainly due to the presence, in the WMAP temperature anisotropy spectrum, of two outliers in correspondence of $\ell = 22$ and $\ell=40$. In particular, these ``glitches'' are well fitted by a primordial power spectrum featuring oscillations localized in 
a suitable range of wave numbers. On the other hand, it is worth noticing that the ``glitches'' could have a more conventional explanation, steaming from some still unknown systematics in the WMAP data.

Features in the primordial power spectrum can be generated following departures from slow roll, that can happen in more general inflationary models. In particular, in multifield supergravity- or M-theory-inspired models \cite{Adams:1996yd, Adams:1997de}, a field coupled to the inflaton can undergo a symmetry-breaking phase transition and acquire a vacuum expectation value. Such a  phase transition corresponds to a sudden change in the inflaton effective mass and can be modeled as a step in the inflationary potential. The presence of the step produces, in turn, a burst of oscillations in the power spectrum of curvature perturbations \cite{Adams:2001vc,Hunt:2004vt}, 
localized around the scale that is crossing the horizon at the time the phase transition occurred. Departures from the standard power-law behaviour can also be present in trans-planckian models \cite{Brandenberger:2000wr,Easther:2002xe,Burgess:2002ub,Martin:2003kp}, in models with a phase of fast roll \cite{Contaldi:2003zv}, or with a sudden change in the speed of sound \cite{Bean:2008na,Nakashima:2010sa,Park:2012rh}. Similarly, in the so-called Starobinsky model \cite{Starobinsky:1992ts}, a change in the slope of the potential causes a step in the perturbation spectrum. In addition to their effect on the power spectrum, these non-standard inflationary scenarios can also be constrained
through their predicted bispectrum \cite{Adshead:2011jq,Martin:2011sn,Park:2012rh}.

The purpose of the present work is to use current data to update previous constraints that have been put on the presence 
of such a step-like feature in the inflaton potential. We improve over previous works by using a more complete dataset 
that includes the WMAP temperature and polarization data, the small-scale CMB data from ACT and SPT, and the matter power spectrum obtained from the Luminous Red Galaxies (LRG) sample of the Sloan Digital Sky Survey (SDSS) 7th data release~\cite{Abazajian:2008wr}). The inclusion of different datasets allows us to explore a wider range of scales with respect 
to previous analyses, going from the Hubble radius down to the smallest linear scales, $k\simeq 0.1$ Mpc$^{-1}$. In particular, this leads to the clear identification of a ``forbidden'' range where oscillations are not allowed.  

The paper is organized as follows: in Section \ref{theory} we briefly recall the theory concerning the evolution of inflationary perturbations in interrupted slow  roll models; in Section \ref{sec:anmeth} we describe the phenomenological model used to describe a step in the inflationary potential, and the analysis method adopted in the present work; in Section \ref{sec:RD} we present the results of the analysis , and in Section \ref{sec:conclusion} we derive our conclusion.

%Constraints on oscillation in the primordial perturbation spectrum, as well as best-fit values for the step parameters,  have been previously derived in Refs. \cite{Peiris:2003ff,Covi:2006ci,Hamann:2007pa,Mortonson:2009qv,Hazra:2010ve}. Here we improve on the previous analyses in several aspects. First, we use more recent CMB data, in particular the WMAP 7-year and the Atacama Cosmology Telescope (ACT) data. This allows us to derive tighter constraints on the parameters; in particular we get an upper limit on the step height (related to the amplitude of oscillations) that is independent on the position of the step itself in the prior range considered. We also find a clear correlation between the position and the height of the step. Secondly, we generate mock data corresponding to the model providing the best-fit to the WMAP data, and use these data to assess the ability of the Planck satellite to detect the presence of oscillations in the primordial spectrum.
%

\section{Inflationary perturbations in models with interrupted slow roll}\label{theory}

\subsection{Inflationary pertubations}\label{how_to}

Let us start by briefly recalling how to compute the spectrum of primordial perturbations for a given inflationary potential $V(\phi)$ \cite{Adams:2001vc}.
In the following we shall work in reduced Planck units ($c=\hbar=8\pi G =1$).
The first step is to solve
the Friedmann and Klein-Gordon equations (dots denote derivatives with respect to the cosmological time $t$):
\begin{align}
3H^2=\frac{\dot\phi^2}{2}+V(\phi) \, ,				\label {eq:Fr}\\
\ddot \phi +3 H \dot\phi + \frac{dV}{d\phi} = 0 \, .  	\label {eq:KG}
\end{align}
to determine the background dynamics of the Hubble parameter $H$ and of the (unperturbed) inflaton field $\phi$.

In order to study the evolution of the curvature perturbation $\mathcal{R}$, one introduces the gauge-invariant quantity
\cite{Sasaki:1986hm,Mukhanov:1988jd,Stewart:1993bc} $u\equiv - z \mathcal{R}$, where $z= a\dot\phi/H$ and $a$ is the scale factor. The Fourier modes $u_k$ of $u$ evolve according to (primes denote derivatives with respect to conformal time $\eta$):
\begin{equation}
u_k''+\left(k^2-\frac{z''}{z}\right) u_k = 0 \, .
\label{eq:u_k}
\end{equation}
In the limit $k^2\gg z''/z$, the solution to the above equation should match the free-field solution $u_k = e^{-i k \eta}/\sqrt{2k}$. The evolution of $z$ is determined directly by the solution of Eqs. (\ref{eq:Fr}) and (\ref{eq:KG}), although during slow roll one can approximate $z''/z \simeq 2a^2 H^2$. At this point, it is possible to integrate Eq. (\ref{eq:u_k}) to get $u_k(\eta)$ for free-field initial conditions.

Finally, the power spectrum of the curvature perturbation $P_\mathcal{R}$ is related to $u$ and $z$ through
\begin{equation}
P_\mathcal{R} = \frac{k^3}{2\pi}\left|\frac{u_k}{z}\right|^2 \, ,
\label{eq:PRk}
\end{equation}
evaluated when the mode crosses the horizon.

\subsection{Models with interrupted slow roll}\label{step_model}

In the following we shall consider models where slow roll is briefly violated. Phenomenologically, these can be described by adding a step feature
to a $V(\phi)=m^2\phi^2/2$ chaotic potential, i.e., by considering a potential of the form 
\begin{equation}
V(\phi) = \frac{1}{2}m^2\phi^2 \left[1+ c\tanh\left(\frac{\phi-b}{d}\right)\right] \, ,
\label{eq:Vstep}
\end{equation}
where $b$ is the value of the field where the step is located, $c$ is the height of the step and $d$ its slope. Although the underlying potential is taken to be the one of chaotic inflation, we shall see below that this form can also be used to describe different kinds of potential.

A sharp step in the inflaton potential, like that described by Eq. (\ref{eq:Vstep}), can appear for example in multi-field inflation models, following a symmetry-breaking phase transition undergone by another field coupled to the inflaton. This induces a rapid variation in the inflaton effective mass $m_\mathrm{eff}$ that is reflected in the potential (indeed, the potential (\ref{eq:Vstep}) is of the form $V(\phi)=\frac{1}{2}m_\mathrm{eff}^2\phi^2$, with a step in $m_\mathrm{eff}$).
In this regard, one can think of $b$ as being related to the time when the phase transition occurs, $c$ to the change in the inflaton mass, and $d$ to the width of the transition.

The spectrum of primordial perturbations resulting from the potential (\ref{eq:Vstep}) can be calculated as outlined in the previous section, and is found to be essentially a power-law with superimposed oscillations. The oscillations are localized only in a limited range of wavenumbers (centered on a value that depends on $b$) so that asymptotically the spectrum recovers the familiar $k^{n_s-1}$ form typical of slow-roll inflationary models. In particular, for a chaotic potential, the underlying power law 
has a spectral index $n_s\simeq 0.96$.

One issue that we have left aside so far is how to relate the horizon size at the time the step occurs to a physical scale. This depends on the number $N_\star$ of e-folds taking place between the time a given mode has left the horizon and the end of inflation.
 We choose $N_\star = 50$ for the pivot wavenumber $k_\star=k_0 =0.05\,\mathrm{Mpc}^{-1}$. This choice is somewhat arbitrary; however, a different choice would correspond to a translation in the position of the step in $\phi$ and would thus be highly degenerate with $b$. For this reason we do not treat $N_\star$ as a free parameter, consistent with what has been done in previous studies \cite{Covi:2006ci,Hamann:2007pa}.

\begin{figure}[t!]   %[1Dlike]
\includegraphics[width=8cm, angle=-90]{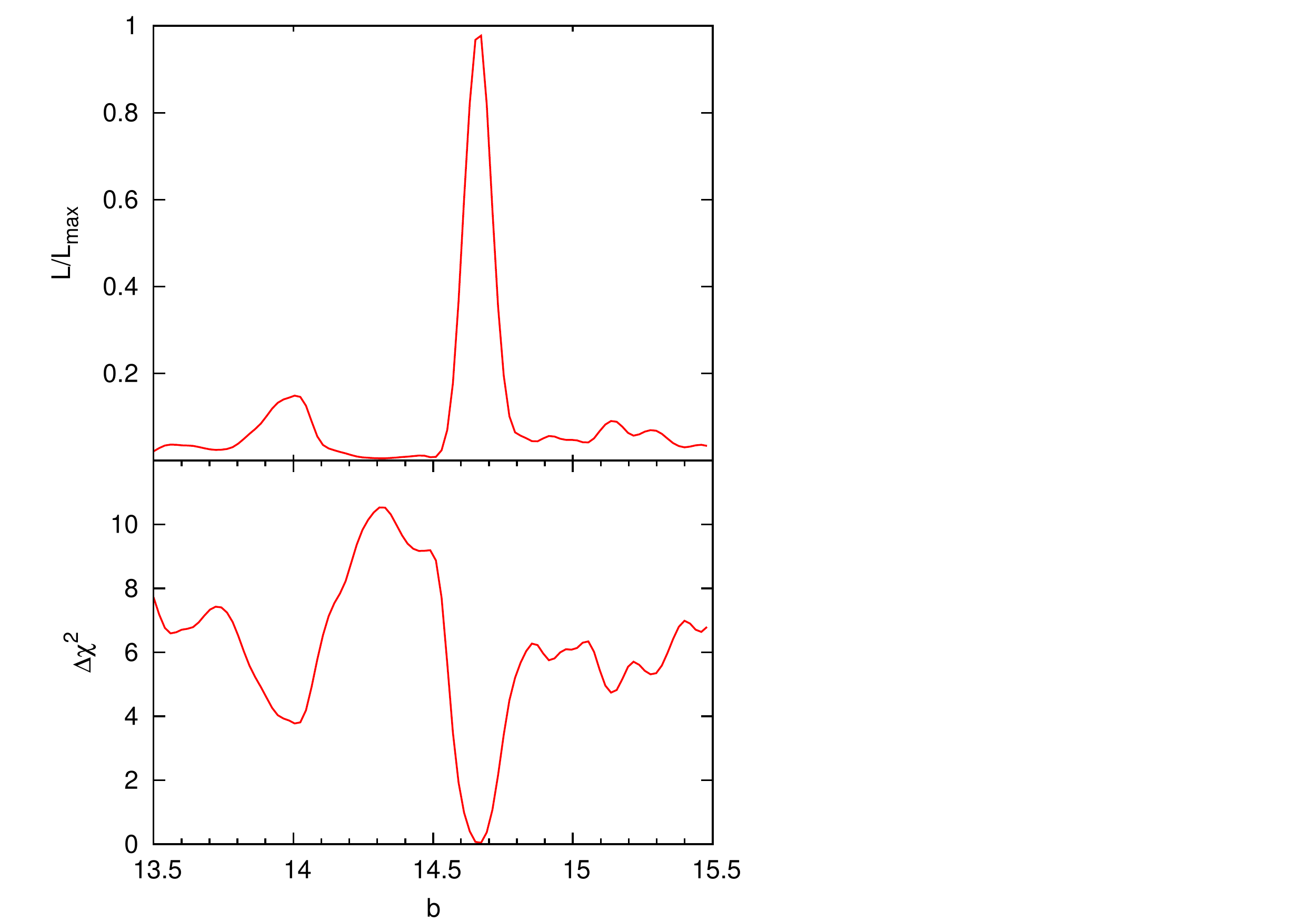}
\caption{Model likelihood (upper panel) and $\Delta\chi^2$ (lower panel) as functions of $b$ for the CMB dataset, obtained by maximization. \label{fig:1Dlike_b_DS1}}
\end{figure}
\begin{figure}
\includegraphics[width=6 cm,angle=-90]{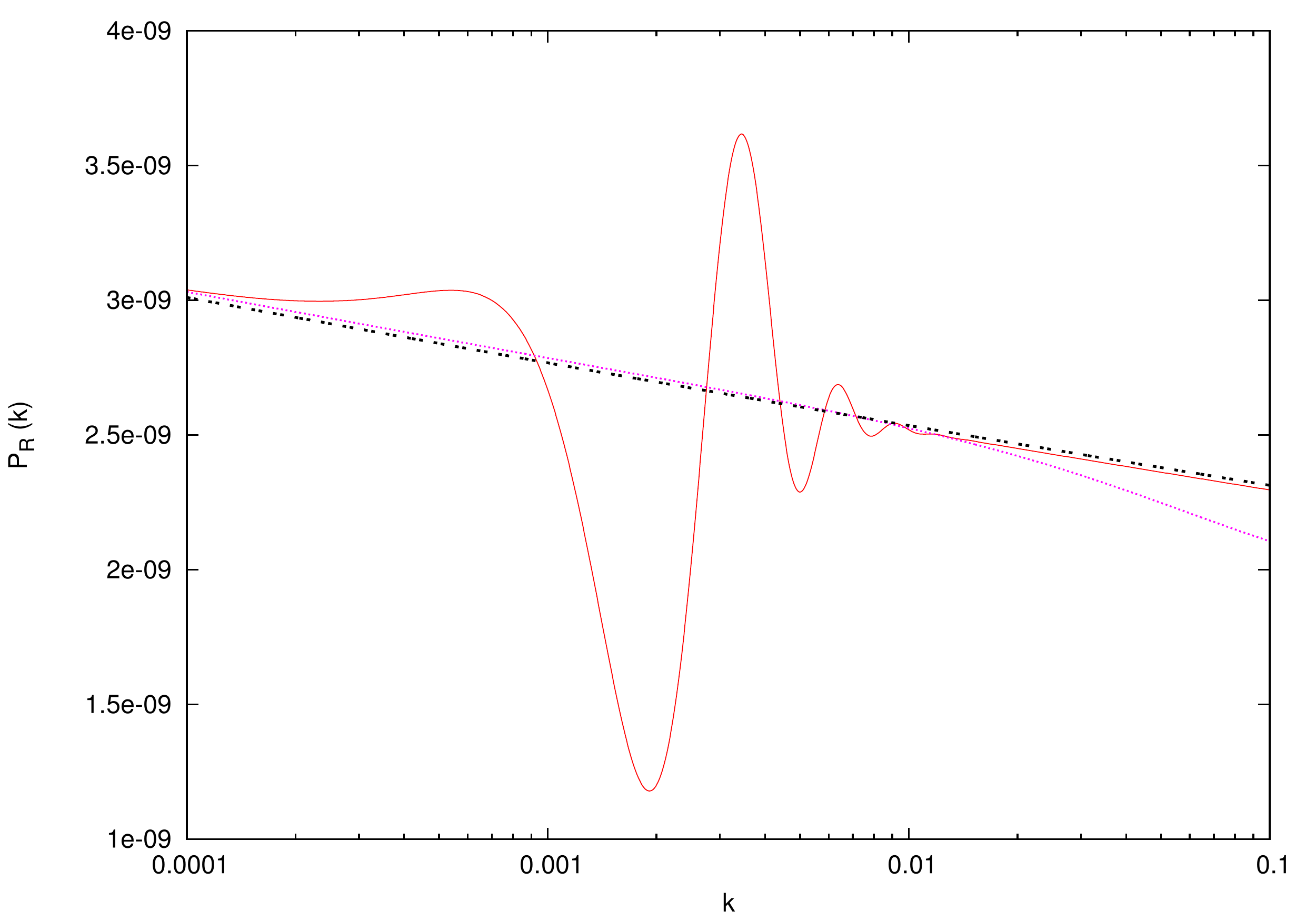}
\caption{Primordial power spectrum for an inflationary potential of the form (\ref{eq:Vstep}) with $m=7.5\times 10^{-6}$. The values of the step parameters are : $b=14.66$, $\log c=-2.75$, $\log d=-1.42$ (red),  $b=14.00$, $\log c=-2.66$, $\log d=-0.54$ (magenta dotted), corresponding to the two peaks in the likelihood. For comparison, we also show the best-fit $\Lambda$CDM power spectrum (black dot-dashed). We note that the model with $b=14$ resembles, in the $k-$range considered, a model with a negative running index.
\label{fig:Pk}}
\end{figure}
\section{Analysis Method \label{sec:anmeth}}

In order to compare the theoretical predictions for the constraints on the parameters characterizing inflationary models with a step in the inflaton potential with observational data,  we performed a Monte Carlo Markov Chain analysis via the publicly available package \verb+CosmoMC+ \cite{Lewis:2002ah}. We used a modified version of the \verb+CAMB+ (\cite{camb}) code in which we numerically solve  Eqs. (\ref{eq:Fr})--(\ref{eq:u_k}) using a Bulirsch-Stoer algorithm in order to theoretically calculate the initial perturbation spectrum (\ref{eq:PRk}), needed to compute the CMB anisotropies spectrum for any given values of the parameters describing this type of inflationary model. Then we compare these theoretical models with two different combination of data sets. We will briefly come back on describing the principal characteristics of each of the dataset considered in this work.

We consider chaotic inflation potentials of the type of Eq. \ref{eq:Vstep}. Following the prescription described in Sec. \ref{how_to}, this potential leads to a well-defined primordial perturbation spectrum $\mathcal{P}_\mathcal{R}$. 
The free parameters in Eq. \ref{eq:Vstep} are then the inflaton mass $m$ and the step parameters $b$, $c$ and $d$.
In our analysis we map the mass $m$ onto $A_s$, i.e. the amplitude of the primordial spectrum at the pivot wavenumber $k_0=0.05\,\mathrm{Mpc}^{-1}$, as indeed the inflaton mass sets the overall scale for the potential and consequently for the amplitude of the perturbations. We note also that the choice of the pivot wavenumber changes the relationship between the value of $b$ and the position of oscillations in $k-$space; this should be taken into account when comparing the results of different studies.
In particular, changing $k_0$ from 0.05 to 0.002 Mpc$^{-1}$ shifts $b$ by $\sim0.5$ towards lower values.

As previously noted in Sec. \ref{step_model}, for the chaotic potential of Eq. \ref{eq:Vstep} the smooth power law has a fixed spectral index, that is  $n_s\simeq 0.96$. However, as noted in Ref. \cite{Hamann:2007pa}, more general forms of the potential can be phenomenologically taken into account by promoting $n_s$ back to a free parameter and defining a ``generalized'' primordial spectrum as 
\begin{equation}
\mathcal{P}^\mathrm{gen}_\mathcal{R}(k) = \mathcal{P}^{\mathrm{ch}}_\mathcal{R}(k)\times \left(\frac{k}{k_0}\right)^{n_s-0.96} \, ,
\label{eq:Pgen}
\end{equation}
where $\mathcal{P}^{\mathrm{ch}}_\mathcal{R}(k)$ is the spectrum induced by the chaotic potential (\ref{eq:Vstep}).

Therefore the theoretical model we are considering is described by the following set of parameters:  

\begin{equation}
 \label{parameter}
      \{\omega_b,\omega_c,\theta,\tau, b, c, d, \mathcal A_s, n_s\}
\end{equation}

where $\omega_b = \Omega_b h^2$ and $\omega_c = \Omega_c h^2$ are the physical baryon and cold dark matter densities, $\theta$ is the ratio between the sound horizon and the angular diameter distance at decoupling, $\tau$ is the optical depth to reionization, $b$, $c$ and $d$ are the parameters of the step-inflation model,  $\mathcal A_s$ is the overall normalization of the primordial power spectrum (equivalent to specifying $m^2$ as discussed above), and $n_s$ is the effective tilt. We consider purely adiabatic initial conditions, impose flatness and neglect neutrino masses, and limit our analysis to scalar perturbations. 

\begin{center}
\begin{figure}   %[disp]
\includegraphics[width=6cm,angle=-90]{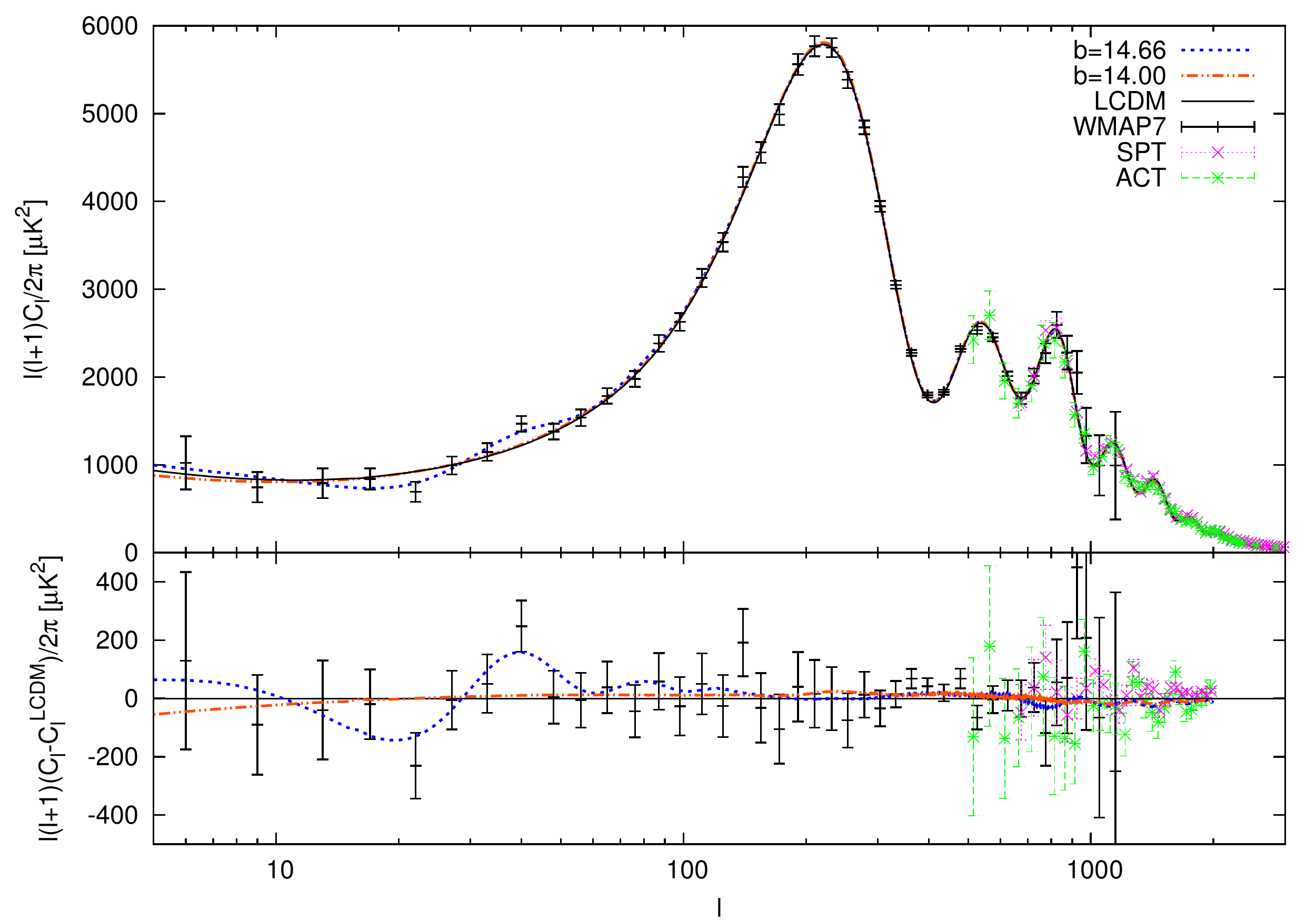}
\caption{Upper panel: Power spectrum of CMB Anisotropies Power Spectrum for the best-fit $\Lambda$CDM model (black line) and two step models with $b=14$ (red dashed) and $b=14.66$ (blue dot-dashed), corresponding to the two minima in $\chi^2$, compared with WMAP7, ACT and SPT data. \
Lower panel: The same as above, but plotted in terms of residuals with respect to the $\Lambda$CDM best fit.
\label{fig:disp_DS1}}
\end{figure}
\end{center}
\begin{center}
\begin{figure}  %[disp_zoom]
\includegraphics[width=6cm,angle=-90]{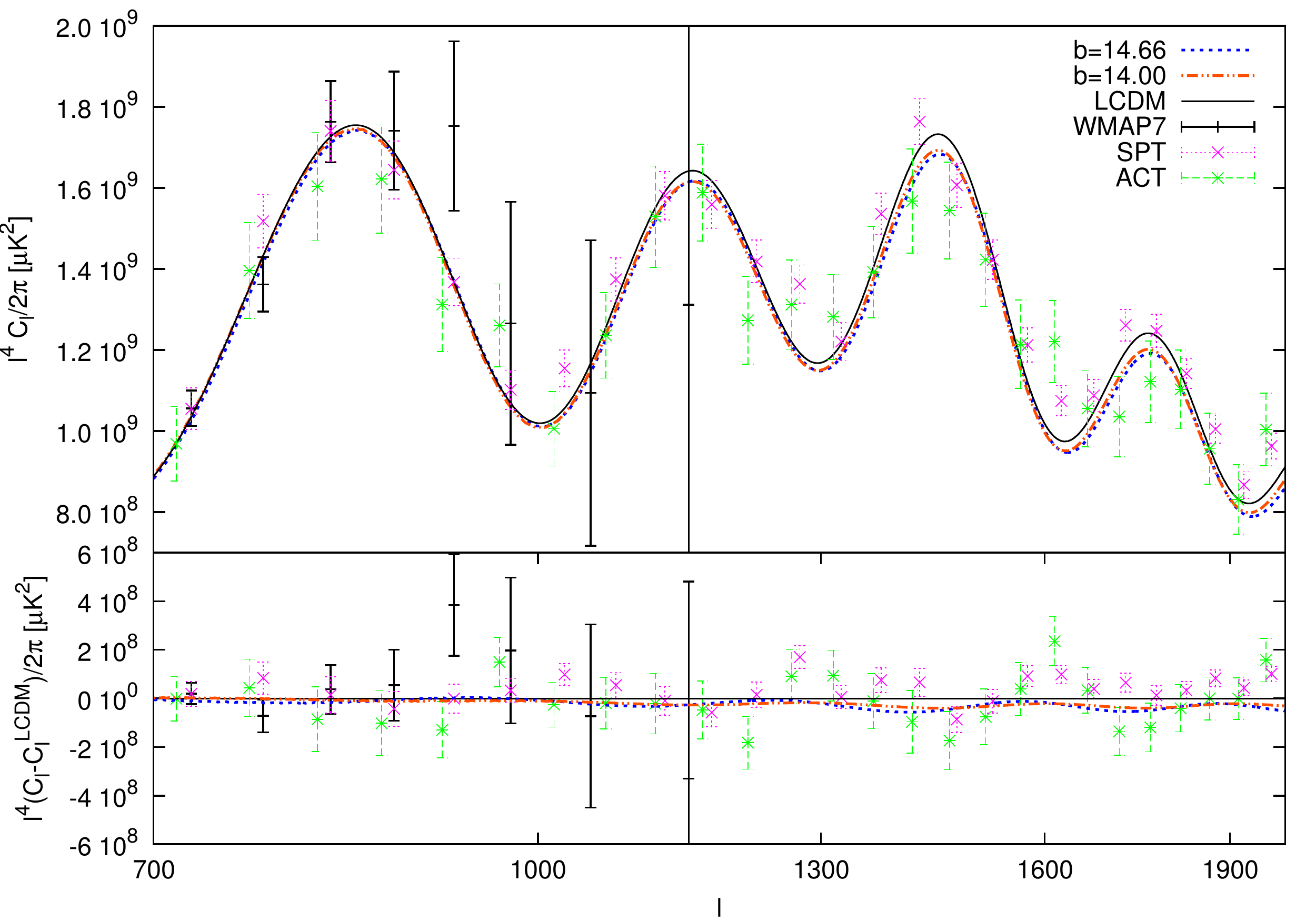}
\caption{The same as Fig. \ref{fig:disp_DS1}, but zoomed in the region $700\le \ell \le 2000$. Note that in order to improve clarity, 
the vertical axis shows $\ell^4 C_\ell$ instead than $\ell(\ell+1) C_\ell$.
\label{fig:disp_zoom_DS1}}
\end{figure}
\end{center}

We consider as our basic dataset a combination of three different CMB datasets: WMAP7, ACT and SPT (in what follows we will refer to this case simply as ``CMB''), and then we also consider an enlarged dataset, dubbed ``CMB+LRG'', where we also add the LRG SDSS catalog \cite{Reid:2009xm}, the Supernovae Ia Union Compilation 2 data \cite{Amanullah:2010vv}, and impose a prior on the Hubble constant from the Hubble Space Telescope (HST) \cite{Riess:2009pu}.

To compute the likelihood of the data we have properly modified the \verb+CosmoMC+ package in oder to make use of the routines supplied by the WMAP and ACT teams for their datasets, both publicly available from the LAMBDA website\footnote{\texttt{http://lambda.gsfc.nasa.gov/}}, and of the likelihood code provided by the SPT team \cite{Keisler:2011aw} for the SPT dataset. 

The ACT and SPT experiments allow to extend the dynamic range of CMB observations to larger multipoles with the respect of WMAP7, thus measuring the damping tail of the CMB angular power spectrum. While SPT probes the small scales in the range of multipoles $650< \ell <3000$, the ACT telescope spans a range of multipoles that goes up to $ \ell=10000$, although the signal at $\ell \gtrsim 3000$ is dominated by the power coming from extragalactic point sources.
For this reason, for ACT we only consider the less contaminated 148 GHz spectrum up to $\ell_{max} =3300$ to perform cosmological parameters extraction. In order to account for the foreground contributions at $\ell\lesssim 3000$, we add three extra amplitude parameters: the Sunyaev-Zel'dovich (SZ) amplitude $A_{SZ}$, the clustered point
sources amplitude $A_C$ and the amplitude of Poisson distributed point sources $A_P$. We consider for both ACT and SPT experiments a joint amplitude parameter for each component and the templates provided by \cite{Keisler:2011aw}. No SZ contribution is considered for WMAP7, as explained in the analysis performed by \cite{Keisler:2011aw}. We have however verified that different choices for the foreground templates has negligible effect on the constraints of cosmological parameters and produces minimal effects on foreground parameters. 

For what concerns the SDSS LRG7 catalog, we chose to consider data only in the linear scales regime, i.e., up to $k=0.1$ h Mpc$^{-1}$. Indeed,  HALOFIT, the \verb+CAMB+ routine that should correct for non-linearity effects at the smallest scales, is tested only for a smooth primordial power spectrum and therefore is not appropriate for dealing with power spectra with features, like those considered in the present analysis.

\begin{table*}[ht!] 
\begin{center}
\caption{Posterior mean for the vanilla cosmological parameters. The errors refer to 68\% credible intervals. \label{tab:Cosmological_posterior}}
\begin{ruledtabular}
\begin{tabular}{cccc}
Parameter		& 
$\Lambda$CDM (WMAP7+CMB)\footnote{Posterior mean for the parameters of the $\Lambda$CDM model, using the WMAP7+CMB dataset from http://lambda.gsfc.nasa.gov.} & 
Features (CMB)\footnote{Posterior mean for the parameters of the features model, using the CMB dataset.} & 
Features (CMB+LRG)\footnote{Posterior mean for the parameters of the features model, using the CMB+LRG dataset} \\
\hline
$100\,\Omega_b h^2$ 		& $2.253\pm0.054$			& $2.204\pm0.044$	 & $ 2.215\pm0.037$	\\
$\Omega_{c} h^2$		& $0.1103\pm0.0052$			& $0.1125\pm0.0050$ & $ 0.1122\pm0.0029$ \\
$100\, \theta$                      & $1.0396\pm0.0025$			& $1.0409\pm0.0016$	 & $ 1.0414 \pm0.0015$	\\
$\tau$ 		                  	& $0.088^{+0.015}_{-0.014}$		& $0.086\pm0.014$	 & $ 0.087 \pm0.015$	\\
$n_s$			  		& $0.962^{+0.014}_{-0.013}$		& $0.959\pm0.014$	& $0.959\pm0.011$	 \\
$10^9 A_s$\footnote{$k_0 = 0.05\,\Mpc^{-1}$. The $\Lambda$CDM value reported in the Lambda website is for $k_0 =0.002\,\Mpc^{-1}$ but has been rescaled
 				to $k_0=0.05$ to allow for comparison.} 
 					& $ 2.15\pm0.11$				& $2.18\pm0.08$  	& $2.19\pm0.07$	 \\
Age [Gyr] 		             	& $13.72\pm0.12$ 			& $13.81\pm0.09$ 	& $ 13.78\pm0.07$ 	\\
$z_{re}$ 		             	& $10.5\pm1.2$				& $10.5 \pm1.2$	& $ 10.5\pm1.2$	\\
$H_0 $ [km s$^{-1}$ Mpc$^{-1}$] 	& $71.4\pm2.4$ 			& $69.9\pm2.3$ 	& $ 70.3\pm1.3$ 
\end{tabular}
\end{ruledtabular}
\end{center}
\end{table*}
\begin{table}
\caption{Bestfit values. \label{tab:Bestfit}}
\begin{ruledtabular}
\begin{tabular*}{0.9\textwidth}{ccc}
Parameter		& {CMB}		&{CMB+LRG}	\\
\hline
$b$ 		                   & $14.66$&	 $14.66$	\\
$\log c$ 		       & $-2.65$	& $-2.85$  \\
$\log d$ 		       & $-1.42$	& $-1.57$	\\
$n_s$			       & $0.946$	& $0.958$  \\
$\ln [10^{10}A_s]$           & $3.08 $	& $3.06$   \\
\hline
$- \log(\mathcal{L})$		       & 3765.8 		& 4043.7  \\
\end{tabular*}
\end{ruledtabular}
\end{table}

Regarding the prior on the model parameters, we impose flat priors on $\omega_b$, $\omega_c$, $\theta$, $\tau$ and $n_s$ and a logarithmic prior on $\mathcal A_s$. 
We check a posteriori that these priors result to be much wider than the corresponding posteriors and thus their upper and lower limits do not affect our final results.
The priors on the step parameters need however to be discussed in more detail. The parameter $b$ controls the position of the oscillations in $k-$space. Larger values of $b$ correspond to ``later'' phase transitions and thus move the oscillations towards larger scales (smaller values of $k$ and $\ell$). Viceversa, smaller values of $b$ shift the oscillations in the direction of large wave numbers. As a rule of thumb, we note that the peak in the oscillations is located at $k\simeq 0.015$ Mpc$^{-1}$ ($\ell \simeq 200$) for $b=14.5$, and that it is shifted down (up) by roughly a factor 2 in $k$ for each 0.1 increment (decrement) in $b$. Thus, outside of a given range in $b$, oscillations are moved to wave numbers that are not probed by observations of the CMB nor of large scale structures. Based on the considerations above, we initially choose a flat prior for $b$ in the range $13.5\le b\le 15.5$, that conservatively encompasses the whole range probed by the WMAP, ACT, SPT and LRG datasets. We use this prior for the CMB only dataset. Then, in view of the results of the first Monte Carlo run, we also consider a restricted prior $14.2\le b \le 15$, that we use for the analysis of the enlarged dataset. We have also explicitly checked that adding, in $b=13.5$ or $b=15.5$, a step-like feature with $c=10^{-2}$ (a value already large enough to produce, on average, oscillations that are at variance with observations \cite{Benetti:2011rp})
and $d=3\times10^{-2}$ (the median point of our prior) to the WMAP7 best-fit model produces no appreciable effect (at least within \verb+CAMB+'s numerical precision) in the CMB spectrum up to $\ell=3000$ nor in the matter power spectrum between $k=0.02$ and $k=0.1$ Mpc$^{-1}$. For what concerns $c$ and $d$, parameterizing the height and width of the step respectively, we choose a logarithmic prior for both of them, i.e., a uniform prior on $\log c$ and $\log d$. The reason for this choice is that we want for these parameters to, potentially, assume values spanning several orders of magnitude with equal \emph{a priori} probability. Indeed this is accomplished using a logarithmic prior that naturally assigns equal probability to each decade. In particular, we take $-4 \le \log c \le -1$ and $-2.5 \le \log d \le -0.5$.

We derive our constraints from parallel chains generated using the Metropolis-Hastings algorithm. We use the Gelman and Rubin $R$ parameter to evaluate the convergence of the chains, demanding that $R-1 < 0.04$. We note that there are some issues related to the fact that some of the posteriors do not vanish at infinity; we address them in section \ref{sec:RD} .

\begin{figure*} %[1Dpost]
\includegraphics[height=0.8\hsize, angle=-90]{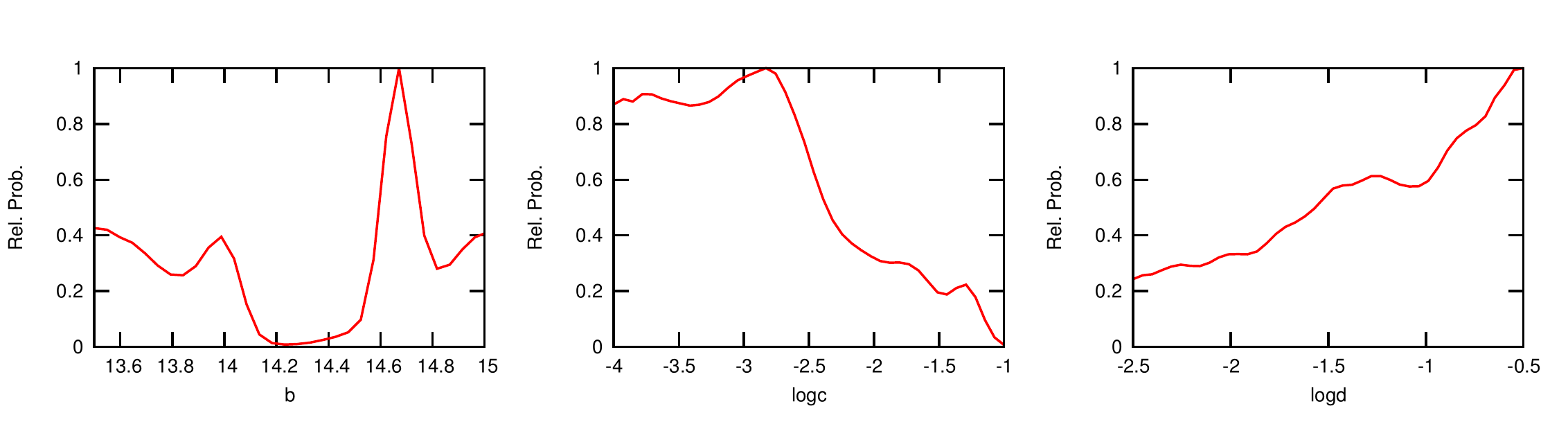}
\caption{One-dimensional posterior probability density for the step parameters from the CMB dataset. \label{fig:1Dpost_DS1}}
\end{figure*}
\begin{center}
\begin{figure}
\includegraphics[width=0.99\hsize]{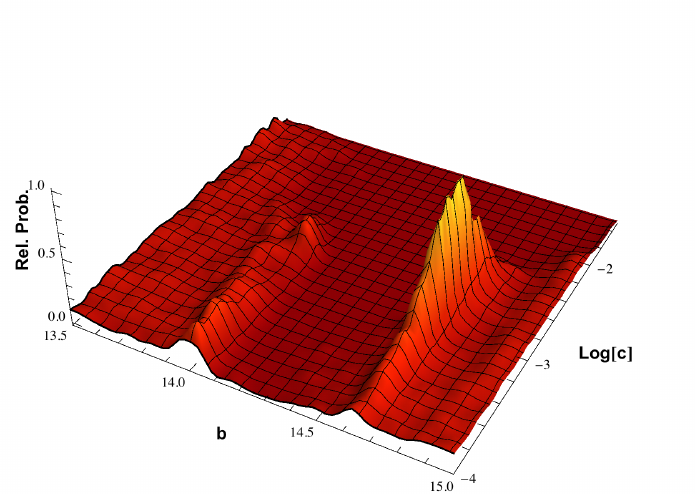}
\caption{Joint two-dimensional posterior for  ($b$, $\log c$) using the CMB dataset. 
\label{fig:2Dpost_DS1}}
\end{figure}
\end{center}

\section{Results and Discussion \label{sec:RD}}

First of all, we check how the constraints on the six ``vanilla'' parameters are changed when the possibility of having features in the primordial power spectrum is considered. To this purpose, we show in Tab. \ref{tab:Cosmological_posterior} the mean of the posterior distribution, and the corresponding 68\% credible intervals, for each vanilla parameter, as well as for some derived parameters (most notably the reionizaion redshift $z_\mathrm{re}$, the age of the Universe, the Hubble constant $H_0$), and compare them with the corresponding values found by the WMAP7 team in their analysis of the $\Lambda$CDM model; however, in order to have the maximum homogeneity between datasets, we consider the dataset dubbed ``WMAP7+CMB'' in the LAMBDA website that includes, in addition to WMAP, also the data from small scale CMB experiments. We note that the uncertainty on the determination of the vanilla parameters is not degraded when features are included (and it is actually better for the CMB+LRG dataset, although this should probably be ascribed to the inclusion of additional data). The mean values found for the features model are all within one sigma of the corresponding $\Lambda$CDM values, with the partial exception of the baryon density $\omega_b h^2$, whose mean lies at the edge of the WMAP 68\% credible interval. We argue that this is due at least in part to the fact that some of the primordial spectra considered here mimic the presence of a negative running in the spectral index (see below); the lower value of $\omega_b h^2$ is thus due to the correlation with the running.
\begin{figure*}[h!]
\includegraphics[width=0.8\hsize]{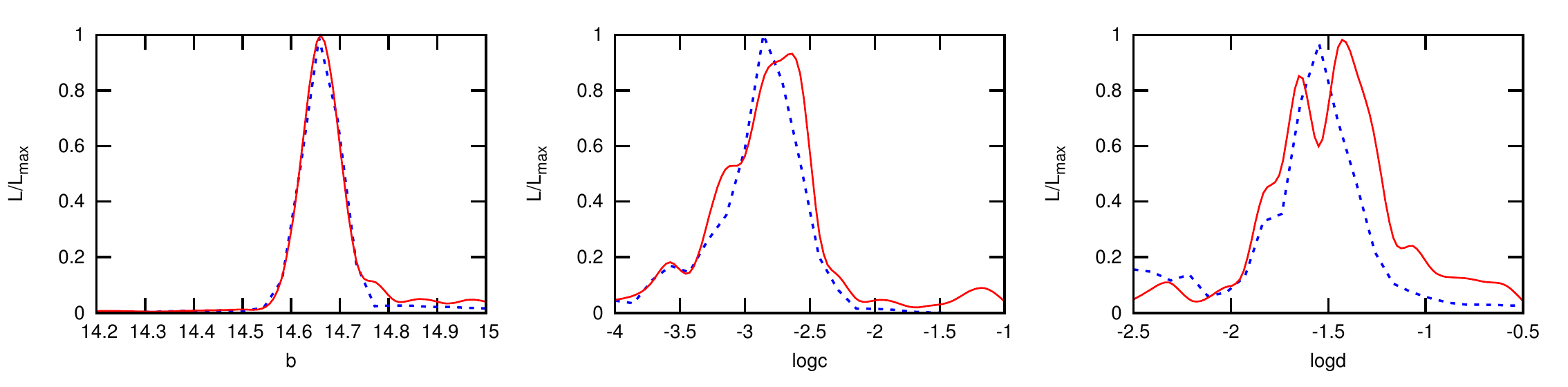}
\caption{Comparison of the model likelihood as a function of the step parameters, obtained from the the CMB (red solid line) and CMB+LRG (blue dashed) datasets. \label{fig:1Dlike_DS12}} 
\end{figure*}
\begin{figure*}[conf_post]
\includegraphics[width=0.8\hsize]{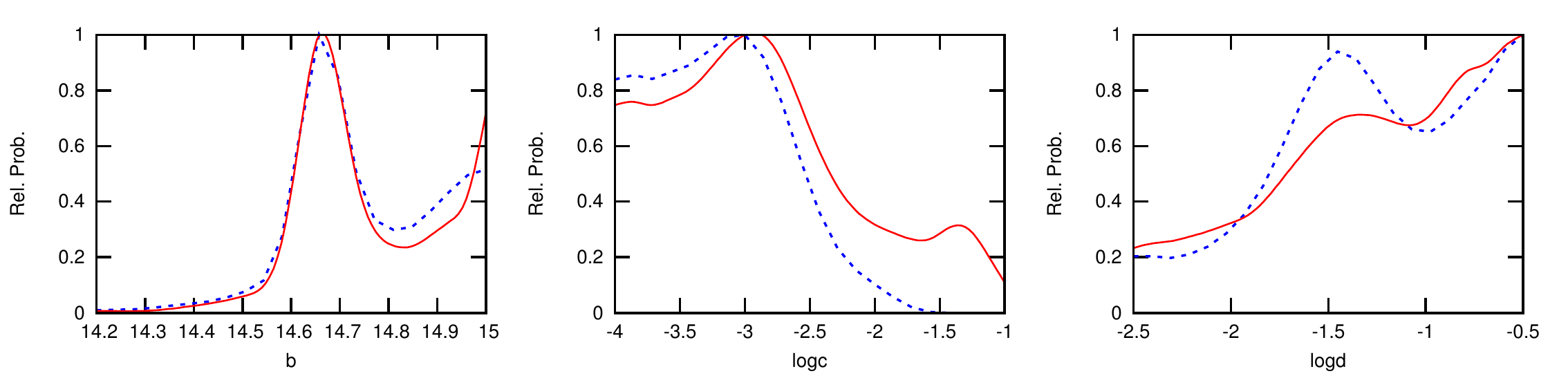}
\caption{Comparison of the one-dimensional posterior probability density for the step parameters from the CMB (red solid line) and CMB+LRG (blue dashed line) datasets.\label{fig:1Dpost_DS12}}
\end{figure*}
Now let us analyze the results on the primordial spectrum parameters from one dataset at a time, starting from the CMB-only dataset. We recall that this analysis assumes the prior $13.5\le b \le 15.5$. We first focus the best-fit parameters, i.e., the parameter values that maximize the likelihood. In the second column of Tab. \ref{tab:Bestfit} we show the best-fit values, for this dataset, of the primordial spectrum parameters. In the best-fit model, the step in the primordial spectrum is located in $b=14.66$, consistently with previous studies \cite{Mortonson:2009qv,Benetti:2011rp}. We explicitly show the projected likelihood, as well as $\Delta\chi^2 = \chi^2-\chi^2_\mathrm{min}$, as a function of $b$ in the two panels of Fig. \ref{fig:1Dlike_b_DS1}. It is interesting however that a distinct although low peak is present in the likelihood in $b\simeq 14$. We also note that the $\chi^2$ does not become arbitrarily large (i.e., the likelihood does not asymptotically vanish) far from the minimum, but instead tends to a constant value. This is related to the fact that, as explained to the previous section, towards the extremes of the $b$ range the oscillations are moved outside the range of scales probed by the dataset, and thus the model becomes completely equivalent, as long as data fitting is concerned, to $\Lambda$CDM. Thus from the plot we can roughly estimate that the best-fit at $b=14.66$ represents a $\Delta \chi^2 \simeq 7$ improvement with respect to the $\Lambda$CDM best-fit, while for the model with $b=14$, $\Delta \chi^2 \simeq 3.5$. The primordial power spectra correspoding to the two minima in the $\chi^2$ are shown in Fig. \ref{fig:Pk}.
On the other hand, models with $14.1\lesssim b \lesssim 14.5$ perform worse with respect to $\Lambda$CDM.

For a better understanding, we also show in Figs. \ref{fig:disp_DS1} and \ref{fig:disp_zoom_DS1} the comparison between the WMAP7 best-fit, the two models with features corresponding to the two peaks in the likelihood seen in Fig. \ref{fig:1Dlike_b_DS1}, and the data present in the CMB dataset. It is clear from these plots (especially from the plot of residuals shown in the lower panel) that the model with $b=14.66$ improves over $\Lambda$CDM by being able to fit the two outliers in $\ell = 22 $ and $\ell = 40$, thus confirming our previous findings \cite{Benetti:2011rp}. The interpretation of the peak in $b=14$ is more puzzling; by looking at the lower panel of Fig. \ref{fig:disp_DS1}, however, it can be seen that the CMB spectrum for this model resembles what it would be obtained by adding a negative running $dn_s/d\ln k$ to the scalar spectral index. Thus this result could be reminiscent of the WMAP7 preference for a negative running, that is indeed even more pronounced when high-$\ell$ data are added to the analysis \cite{Komatsu:2010fb}: $dn_s/d\ln k= -0.034\pm0.026$ (WMAP7 only) and $dn_s/d\ln k= -0.041^{+0.022}_{-0.023}$ (WMAP7+ACBAR+QUaD).

We now turn to the posterior distributions. In Fig. \ref{fig:1Dpost_DS1}, we show the one-dimensional posteriors for $b$, $\log c$ and $\log d$. The posterior for $b$ still shows the two peaks in $b\simeq 14.7$ and $b\simeq 14$ that were present in the likelihood. The largest value at the edges of the prior range is due to a volume effect, since the one-dimensional posterior is obtained by marginalization (as opposed to the one-dimensional likelihood that was obtained by maximization). On the other hand, the probability density for $14.1\lesssim b \lesssim 14.4$ is practically equal to zero. For what concerns $\log c$, as it could be expected, ``large'' values are disfavored by the data (as they produce large - in amplitude - oscillations that cannot, on average, be reconciled with observations) while for smaller values the posterior tends to a constant value as the oscillations become so weak as to be practically undetectable for the current experimental precision and thus the value of $c$ becomes unimportant. As already noted, a posterior with this characteristic that extends, in principle, down to $\log c = -\infty$, cannot be properly normalized (since the corresponding probability mass is infinite) and, as a consequence, credible intervals are ill-defined. One could be tempted to impose a lower cut-off but then the credible intervals will end up depending on the choice of the cut-off itself, so this should be avoided, at least in the absence of a clear physical reason for doing so. 

We can still, however, compare probability densities, as well as probabilities integrated over finite intervals, since probability ratios \emph{do not} depend on the overall normalization. We can use, as a benchmark value to compare the constraining power of different datasets, for example, the value of $\log c$ where $P(\log c )$ is half of its asymptotic value for $\log c\to \infty$. This should not be taken as an ``upper limit'' in the common sense of the word, but as said is a useful tool for comparison. In the case under consideration, we estimate that this happens for $\log c =-2.32$, or $c = 4.8 \times 10^{-3}$. 

For comparison, the corresponding value that we had previously found using WMAP7 and ACT data only was $\log c = -2$ \cite{Benetti:2011rp}. We also show, in Fig. \ref{fig:2Dpost_DS1} the two-dimensional posterior $P(b,\, \log c)$ where it is clear that probability is concentrated in two distinct, disconnected regions. One corresponds to models with $b \simeq 14.7$ and $\log c \simeq -3$, while the other to models with $b \simeq 14$ and $\log c $ located more towards the edge of the prior range, $\log c\lesssim -3.5$. Finally, we examine the posterior for $\log d$. This is in part similar to the posterior for $\log c$, once one recalls that small values of $\log d$ produce a steep step in the potential and consequently large oscillations, so one should expect the probability to go to zero for small values of $\log d$, as it is. However, in this case, the posterior range is not wide enough to see the asymptotic part, for $\log d \to \infty$ (where $\Lambda$CDM should be recovered), of the distribution.

The fact that the posterior is bimodal in $b$ creates some difficulty for the Monte Carlo, as the chains cannot easily jump from one peak to the other, and thus
take a longer time to sample satisfactorily the actual distribution. For this reason, in our second Monte Carlo run, using the CMB+LRG dataset, we have decided to concentrate on the region of the peak at $b=14.66$ and impose the prior $14.2\le b \le 15$. We find that the best-fit for this dataset, shown in the third column of Tab. \ref{tab:Bestfit} has still $b=14.66$. In Figs. \ref{fig:1Dlike_DS12} and \ref{fig:1Dpost_DS12} we compare the one-dimensional likelihoods and posteriors, respectively, for the step parameters in the CMB+LRG dataset with those obtained previously with the CMB dataset. In order to allow for comparison, the distributions for the latter have been obtained by imposing a posteriori the condition $b\ge 14.2$ (which, in practical terms, that we have discarded all samples with $b<14.2$, and reanalyzed these new chains from scratch). We find that there is practically no difference with respect to the position of the oscillations (which makes sense, since this is driven by the requirement of fitting the outliers in the WMAP7 data at relatively low $\ell$'s). The amplitude of the oscillations is slightly more constrained, with the posterior going down at half of its plateau value at $\log c = - 2.48$ ($c\simeq 3\times 10^{-3}$). The posterior for $\log d$ is also slightly different, as it shows a more distinct peak in correspondence of the best-fit value $\log d\simeq -1.5$.

\section{Concluding Remarks}\label{sec:conclusion}

We have studied cosmological models with a step-like feature in the inflationary potential. Such a feature would produce oscillations in the primordial spectrum of scalar perturbations, whose presence can be tested through the analysis of CMB and large-scale structures data. We have found, consistently with previous studies, that in these models the agreement with the CMB data is improved, with respect to the $\Lambda$CDM model, when the oscillations are placed in such a way as to match the two outliers in the WMAP7 spectrum at $\ell=22$ and $\ell=40$ (in particular, the $\chi^2$ changes by $\Delta\chi^2 \simeq 7$). The posterior probability also has a maximum close to this point, corresponding to $b=14.66$, while it clearly shows that oscillations in the range $14.1\le b \le14.5$ are currently forbidden by the data. The possibility of no oscillation at all is still, however, perfectly consistent with the data. In conclusion, although multifield inflationary models can definitely reproduce the two glitches in the WMAP7 temperature spectrum, current data are not yet constraining enough to allow to discriminate between these models and the standard inflationary scenario.

\section{Acknowledgments}
The Dark Cosmology Centre is funded by the Danish National Research Foundation. The work of ML has been supported by Ministero dell'Istruzione, dell'Universit\`a e 
della Ricerca (MIUR) through the PRIN grants ``Matter-antimatter asymmetry, Dark Matter and Dark Energy in the LHC era'' (contract number PRIN 2008NR3EBK-005) and ``Galactic and extragalactic polarized microwave emission'' (contract number PRIN 2009XZ54H2-002). We would like to thank Erminia Calabrese and Luca Pagano for useful discussion.

%

%\newpage

\end{document}